\documentclass[showpacs,prd,eqsecnum]{revtex4}
\usepackage{amsmath,amsthm,amssymb}
\usepackage{graphics}
\usepackage{color}
\usepackage{dcolumn}
\begin{document}
\preprint{}
\title{Observational constraints of a power spectrum from super-inflation in Loop Quantum Cosmology}
\author{Masahiro Shimano}
 \email{shimano@rikkyo.ac.jp}
\author{Tomohiro Harada}
  \email{harada@rikkyo.ac.jp}
\affiliation{Department of Physics, Rikkyo University, Toshima, Tokyo 171-8501, Japan}
\date{\today}
\begin{abstract}
In loop quantum cosmology there may be 
a super-inflation phase in the very early universe, 
in which a single scalar field with a negative power-law potential 
$V= -M^4\left(\phi/M\right)^\beta$ plays important roles. 
Since the effective horizon $\sqrt{SD}/H$ controls 
the behavior of quantum fluctuation instead of the usual Hubble horizon, 
we assume the following inflation scenario; the super-inflation starts 
when the quantum state of the scalar field emerges into the 
classical regime,
and ends when the effective horizon becomes the Hubble horizon, 
and the effective horizon scale never gets 
shorter than the Planck length. 
From consistency with the WMAP 5-year data, 
we place a constraint on the parameters of the potential 
($\beta$ and $M$) and the energy density at the end of the super-inflation, 
depending on the volume correction parameter $n$. 
\end{abstract}

\maketitle

\section{Introduction}
In the end of the previous century, loop quantum gravity (LQG)~\cite{c0,t2,Ashtekar_2005}
inspired loop quantum cosmology (LQC), 
which is an application of loop quantization to the 
homogeneous universe models~\cite{b0,b1,b2,b3,bojo0,bojo1,bojo2}. This is featured with 
singularity avoidance and super-inflation, where  
the Hubble parameter increases with time. 
The semiclassical effects of LQG can be
incorporated in the forms of the volume correction~\cite{bojo1,bojo2} 
and the energy density correction coming from the holonomy 
effects~\cite{Date:2004,Banerjee:2005,Singh:2005,As1,As2,As3} 
into the Hamiltonian.

The observation of anisotropy in the cosmic microwave background (CMB)
will be the most powerful tool available at present and in the near future 
to probe the inflationary phase of the universe. 
It is an interesting possibility that 
the loop quantum effects in the very early phase of the universe 
might be imprinted in the CMB anisotropy.
In this context, Tsujikawa et al.~\cite{tsuji}
showed that the dynamics during super-inflation due to the volume
correction can drive an inflaton 
field up its potential hill, thus setting the initial conditions 
for the standard slow-roll inflation and suggested that this transition 
from the super-inflation to the standard inflation might be responsible 
for the observed loss of power at the largest angles in the 
CMB power spectrum but without any explicit calculation of
quantum fluctuation in the super-inflationary phase.
On the other hand, Zhang and Ling~\cite{Xin} considered the 
slow-roll super-inflation 
phase due to the energy density correction and calculated quantum 
fluctuation of the inflaton but without the volume correction effect. 
They found that signature of loop quantum effects is too weak
to detect in the CMB power spectrum with reasonable sensitivity. 

If there is a non slow-roll 
super-inflationary phase due to the loop quantum effects,
we can infer that quantum fluctuation generated in that phase might
leave imprints in the primordial density perturbation because
the statistical properties, say non Gaussianity, 
of quantum fluctuation generated in 
the super-inflation would be sufficiently different from that 
generated in the standard slow-roll inflation.
However, for such a scenario
to be viable, the predicted power spectrum of the density perturbation 
must be sufficiently scale-invariant as observed now in the 
CMB power spectrum and large scale structure. 
Mulryne and Nunes~\cite{M1} investigated this issue with a scalar field 
with a power-law potential $V\propto \phi^\beta$,
for which the volume correction is incorporated into the Hamiltonian.
They showed that for a non slow-roll solution with constant ratio between the kinetic and potential energies of the scalar field, which is called a scaling solution, 
if we take the limit $\beta\to \infty$,  
the density perturbation generated in the super-inflationary phase is scale-invariant.
Copeland et al.~\cite{M2} showed that the scaling solution corresponds to a 
stable fixed point in terms of dynamical systems theory and that 
the potential must be negative.

However, taking the limit $\beta\to \infty$ is not physically acceptable
and in reality it is important to determine the allowed region of the 
parameter(s) for the scenario to be consistent with the 
presently available observational data, 
in particular the CMB power spectrum. 
In this paper,  
we focus on the following situation according to Refs~\cite{M1,M2}: 
$\mathrm{i}$) the dynamics is affected from the volume correction
both in the matter and the gravitational Hamiltonians~\cite{V1},
$\mathrm{ii}$) the scalar field has a negative potential and 
$\mathrm{iii}$) its evolution follows a scaling solution.
We adopt the following inflation scenario:
the super-inflation starts when 
the scalar field emerges into the classical regime from the quantum regime, 
and ends when the semiclassical corrections from loop quantum effects
become insignificant. 
We require that the calculated power spectrum is consistent with 
the WMAP 5-year data~\cite{k} and 
obtain the allowed region for the parameters of the scenario.
It is interesting that we can put an upper bound as well as 
a lower bound on $\beta$ and 
very large values of $\beta$ are 
disfavored because of the observed significant deviation 
from the scale-invariant power spectrum.

This paper is organized as follows.
In section $\mathrm{II}$ we review Refs~\cite{M1,M2}.
In section $\mathrm{III}$ we place a constraint on the parameters, 
and Section $\mathrm{IV}$ is a conclusion.
In Appendix A we review the calculation of the power spectrum 
of quantum fluctuation
given by Ref~\cite{M2} with volume correction in the gravitational Hamiltonian. 
In this paper we use the units in which $c=\hbar=1$.
\section{The quantum fluctuation in LQC}
We here review the derivation of the power spectrum and the scaling
solution according to Refs~\cite{M1,M2}, 
but with the other volume correction to the Hamiltonian~\cite{V1}.
\subsection{The loop quantized Hamiltonian}\label{section2}
We first consider the homogeneous and isotropic 
universe described by the FRW metric 
       	\begin{equation}
	 ds^2=-dt^2+a(t)^2\left[dr^2+r^2\left(d\theta^2+\sin^2\theta d\phi^2 \right)\right]\label{a}, 
	\end{equation}
where $a$ is the scale factor.
In LQC the Hamiltonian for gravitation and a single scalar field
is given by $\mathcal{H}=\mathcal{H}_{\rm grav}+\mathcal{H}_{\rm matter}$, where\cite{bojo3,V1} :
	\begin{eqnarray}
	 \mathcal{H}_{\rm grav}&=&-\frac{3}{8\pi\gamma G}aS(q)\dot{a}^2, 
\label{gravhamil} \\
	 \mathcal{H}_{\rm matter}&=&\frac{D(q)}{2a^3}p_{\phi}^2+a^3 V(\phi)\label{mattehamil}.
	\end{eqnarray}
$S(q)$ and $D(q)$ are respectively the volume correction 
factors in the gravitational and the scalar field Hamiltonians, 
and the dot denotes the derivative with respect to $t$. 
We first assume that 
the scalar field depends only on time, i.e. $\phi=\phi(t)$.
$p_{\phi}$ is a conjugate momentum of $\phi$, which is defined as
	\begin{equation}
	 p_{\phi}=-\frac{a^3\dot{\phi}}{D(q)}\label{conjugatemoment},
	\end{equation}
and $q$ is defined as \cite{As1,As2,As3}: 
	\begin{equation}
	 q\equiv\left(\frac{a}{a_*}\right)^3,
\label{eq:q}
\end{equation}
where 
\begin{equation}
a_*=\left(\frac{2j}{K}\right)^{\frac{1}{3}}\sqrt{\frac{4\pi\gamma}{3}}l_{\mathrm{Pl}},\quad K=\frac{2\sqrt{2}}{3\sqrt{3\sqrt{3}}},\quad l_{\mathrm{Pl}}^2=G \label{a_*},
\end{equation}
$j$ is an SU(2) parameter which is associated with the link of 
the spin network state in LQG (we assume $j$ is sufficiently large), 
and $\gamma$ is the Barbero-Immirzi parameter which is here 
assumed $\gamma=\ln(2)/(\pi\sqrt{3})$ by 
the black hole entropy argument in LQG \cite{c0},
but see also Ref~\cite{tamaki_nomura_2005}.
Note that $a_*$ is the characteristic scale factor in LQC: when the scale factor is smaller than $a_*$, the LQC effects are remarkable.
In the semiclassical region ($l_{\mathrm{Pl}}\ll a\ll a_*$), 
$S(q)$ and $D(q)$ take the following forms~\cite{s1}:
	\begin{equation}
	 S(q)\sim \frac{3}{2}q,\quad D(q)\sim\left(\frac{9}{2l+3}\right)^{\frac{3}{2(1-l)}}q^{\frac{3(3-l)}{1-l}}\label{s0},
	\end{equation}
while, in the classical region ($a\gg a_*$), these are
	\begin{equation}
	 S(q)\sim 1,\quad D(q)\sim 1, 
	\end{equation}
so that the classical theory is recovered. 
It should be noted that the characteristic scale factor $a_*$
is in fact a problematic object in LQC as indicated in Refs~\cite{G.Calcagni1, G.Calcagni2}.
We will caution this in section $\mathrm{IV}$.
We can conveniently parametrize the correction factors 
in the semiclassical region as follows:
	\begin{equation}
	 S(q)=S_* a^r,\quad D(q)=D_* a^n,\label{correctionterm01}
	\end{equation}
where 
	\begin{eqnarray}
	 S_*&=&\frac{3}{2}a_*^{-r}\label{sstar}, \\
 	 D_*&=&\left(\frac{9(9-n)}{81-5n}\right)^{\frac{n-9}{12}}a_*^{-n}\label{dstar},\\
	 n&=& \frac{9(3-l)}{1-l},\\
	 r&=&3. 
	\end{eqnarray}
We assume $0<l<1$ and hence $9<n<\infty$ 
to remove the divergence of the inverse volume factor.
Notice that 
for Eq.~(\ref{dstar}) to be physical, 
we impose the constraint  $81/5<n<\infty$ on the parameter $n$.

The scalar field in general depends on time and position, 
i.e. $\phi=\phi(t,x)$. So in this situation 
we need to consider the following scalar field Hamiltonian in LQC:
	\begin{equation}
	 \mathcal{H}_{\rm matter}=\frac{D(q)}{2a^3}p_{\phi}^2+
aS(q)\delta^{ab}
\partial_a\phi\partial_b\phi+a^3 V(\phi),\label{materhamiltonian}
	\end{equation}
where $a$ and $b$ run over all spatial induces or $1$ to $3$.
We will consider quantum fluctuation of the scalar field for this Hamiltonian.
From the gravitational and the scalar field Hamiltonians~(\ref{gravhamil})
and (\ref{mattehamil}), we can obtain 
the following modified Friedmann equation and the scalar field equation: 
	\begin{eqnarray}
	&& H^2 = \frac{8\pi G}{3}S(q)\rho \label{f0}, \\
	&& \ddot{\phi}+3H \left( 1-\frac{1}{3}\frac{d\ln D}{d\ln a} \right) \dot{\phi} + DV_{,\phi}=0\label{e0},
	\end{eqnarray}
where $_{,\phi}$ denotes the derivative with respect to $\phi$ 
and $H\equiv \dot{a}/a$ is the Hubble parameter, and 
the energy density $\rho$ is given by
	\begin{equation}
	 \rho=\frac{\dot{\phi}^2}{2D(q)} + V(\phi).\label{fooo}
        \end{equation}
Using the conformal time $\tau$ where $dt=ad\tau$, 
Eq.~(\ref{e0}) is rewritten as
	\begin{equation}
	 \phi''+\left(2-\frac{d\ln{D}}{d\ln{a}}\right)\frac{a'}{a}\phi'+a^2DV_{,\phi}=0\label{e1},
	\end{equation}
where the prime denotes the derivative with respect 
to $\tau$. 
In the classical region, since $D(q)\simeq S(q)\simeq 1$, these equations
reduce to the classical ones, 
on the other hand, in the semiclassical region, 
the second term on the left-hand side of Eq.~(\ref{e1}) acts an
anti-friction term and enables the scalar field to climb the potential.

More remarkably, the time derivative of the Hubble parameter
	\begin{equation}
	 \dot{H}=-\frac{4\pi GS\dot{\phi}^2}{D}\left[1-\left( \frac{1}{6}\frac{d\ln{D}}{d\ln{a}}+\frac{1}{6}\frac{d\ln{S}}{d\ln{a}} \right) \right]+\frac{4\pi GS}{3}\frac{d\ln{S}}{d\ln{a}}V
	\end{equation}
is positive in the semiclassical region. The 
accelerated expansion with this feature
is called super-inflation.

Since we have $c=\hbar=1$, the length, the time and 
the mass are all of the same dimension, say $L$.
The dimensions of the scale factor $a$ and the Newtonian constant $G$ 
are then $L$ and $L^2$. 
From this argument, Eq.~(\ref{f0}) shows that the scalar field $\phi$ 
and its potential $V$ are of dimension $L^{-1}$ and $L^{-4}$, respectively.
\subsection{The scaling solution}
We review the scaling solution and its stability according to 
Refs~\cite{M1,M2}. 
We write the dynamics of the homogeneous system in terms of the following 
three variables:
	\begin{equation}
 	 x = \frac{\dot{\phi}}{\sqrt{2D \rho}} , \quad  y = \frac{\sqrt{\vert V \vert}}{\sqrt{\rho}}, \quad  \lambda = -\sqrt{\frac{3D}{16\pi GS}}\frac{V_{,\phi}}{V}\label{def},
 	\end{equation}
where we use Eq.~(\ref{fooo}) with the negative potential, and we also 
need a constraint 
	\begin{equation}
	 x^2- y^2=1\label{condition}
	\end{equation}
from the Friedmann equation~(\ref{f0}). 
From Eqs.~(\ref{f0}), (\ref{e0}) and (\ref{def}), we can derive the following set of ordinary differential equations:
	\begin{eqnarray}
	 x,_N&=&3x\bar{\alpha} - 3 x^3\bar{\alpha} + \sqrt{\frac{3}{2}} \lambda y^2 \label{x0}, \\
	 y,_N&=&-3\bar{\alpha} y x^2-\sqrt{\frac{3}{2}}\lambda xy \label{y0}, \\
	 \lambda,_N&=& \frac{\lambda}{2}(n-r)+\sqrt{6}x\lambda^2\left(1-\Gamma \right) \label{z0},
	\end{eqnarray}
where $N\equiv \ln a$ and
	\begin{equation}
	 \bar{\alpha}=\frac{n}{6}-1,\quad\Gamma=\frac{V V_{,\phi\phi}}{V_{,\phi}^2}\label{gamma00}.
	\end{equation}
Because of the constraint (\ref{condition}), 
$x$ and $y$ are not independent.
Hence we consider only $x$ and $\lambda$. 
This system has several fixed points but here we concentrate on stable
ones only. If
	\begin{equation}
	 \Gamma<\frac{n+r-12}{2(n-6)},
	\end{equation}
a couple of stable fixed points $(x,\lambda)$ are given by 
        \begin{equation}
	 \left(1,-\frac{n-r}{2\sqrt{6}(1-\Gamma)}\right), \quad 
\left(-1,\frac{n-r}{2\sqrt{6}(1-\Gamma)}\right)\label{koteiten0}.
	\end{equation}
These points correspond to the kinetic-term dominant solutions. 
If
	\begin{equation}
	 \frac{n+r-12}{2(n-6)}<\Gamma<\frac{12+n-3r}{2(6-r)},
	\end{equation}
a couple of stable fixed points are given by
        \begin{equation}
	 \left(-\sqrt{\frac{n-r}{12\bar{\alpha}(1-\Gamma)}},
 \sqrt{\frac{\bar{\alpha}(n-r)}{2(1-\Gamma)}} \right),
\quad \left(\sqrt{\frac{n-r}{12\bar{\alpha}(1-\Gamma)}},-\sqrt{\frac{\bar{\alpha}(n-r)}{2(1-\Gamma)}} \right)\label{koteiten1}.
	\end{equation}
These points correspond to the scaling solutions where 
the ratio of the kinetic term to the potential term is kept constant.
For simplicity, we only consider a constant $\Gamma$ and 
then we can determine the potential form by Eq.~(\ref{gamma00}).
For $\Gamma\neq 1$, the potential is given by 
	\begin{equation}
	 V=-V_0 |\phi|^\beta\label{plow},
	\end{equation}
where $\beta$ and $V_0$ are constants.
For $\Gamma=1$, we have an exponential potential.
Notice that since the kinetic-term dominated solution (\ref{koteiten0}) 
and the scaling solution (\ref{koteiten1}) can not be defined for $\Gamma=1$,
we can only use the power-law potential (\ref{plow}) and 
have $1-\Gamma=1/\beta$.
Moreover, since the scaling solutions can be
responsible for the fluctuation of the present CMB radiation as we will see later,
we hereafter adopt the scaling solutions.

To calculate quantum fluctuation we rewrite Eq.~(\ref{def}) as  
	\begin{equation}
	 x=\sqrt{\frac{4\pi G}{3}\frac{S}{D}}\phi_{,N}\label{teigi0}.
	\end{equation}
The two scaling solutions expressed by fixed points
(\ref{koteiten1}) can be analyzed in the same manner,
we take the first one
	\begin{equation}
	 x_0=-\sqrt{\frac{(n-r)\beta}{2(n-6)}}\label{xxxo},
	\end{equation}
where we have used Eqs.~(\ref{gamma00}) and (\ref{plow}). 
Substituting $x=x_0$ into Eq.~(\ref{teigi0}), and integrating 
it with respect to $a$,
we obtain $\phi$ as
 	\begin{equation}
	 \phi=
         \frac{2x_0}{n-r}\sqrt{\frac{3D}{4\pi GS}}\label{phiphi}.
	\end{equation}
Differentiating the above with respect to $\tau$, we obtain
	\begin{equation}
	 \phi'=
         x_0\sqrt{\frac{3D}{4\pi GS}}\frac{a'}{a}\label{phiphi1}.
	\end{equation}
On the other hand, we rewrite Eq.~(\ref{def}) as 
	\begin{equation}
	 \phi'=
         x_0a\sqrt{2D\rho}\label{phiphi0}. 
	\end{equation}
Here we can write $\rho$ in terms of $x_0$ using Eqs.~(\ref{def}) and (\ref{condition}).
For consistency between Eqs.~(\ref{phiphi1}) and (\ref{phiphi0}), 
we can obtain the differential equation of the scale factor. 
Integrating it with respect to $\tau$, 
the scale factor is obtained as
	\begin{equation}
	 a=A(-\tau)^p\label{acalefactor},
	\end{equation}
where 
	\begin{eqnarray}
	 A&=&\left[-\frac{1}{p}\sqrt{\frac{8\pi f\tilde{S}_*}{3(x_0^2-1)}}\left|\frac{2x_0}{n-r}\sqrt{\frac{3\tilde{D}_*}{4\pi \tilde{S}_*}}\right|^{\frac{\beta}{2}}\right]^p\left[\left(\frac{2j}{K}\right)^{\frac{1}{3}}\sqrt{\frac{4\pi\gamma}{3}}\right]^{p+1}l_{\mathrm{Pl}}\label{jigen1},\\
	 p&=&-\frac{4}{2(r+2)+(n-3)\beta}\label{poten0},\\
	 \tilde{S}_*&=&\frac{3}{2},\\
         \tilde{D}_*&=&\left(\frac{9(n-9)}{5n-81}\right)^{\frac{n-9}{12}}\label{DefD}.
	\end{eqnarray}
Since $\phi$ and $V_0$ have dimension 
of $L^{-1}$ and $L^{\beta-4}$, respectively, we put this $V_0$ as
	\begin{equation}
	 V_{0}=\frac{f}{l_{\mathrm{Pl}}^{4-\beta}}=M^{4-\beta},\label{defM}
	\end{equation}
where $f$ is a dimensionless constant and $M$ gives the 
mass scale of the scalar field $\phi$.
We have used this form to get Eq.~(\ref{jigen1}).
\subsection{The effective horizon}

In the usual inflation scenario, quantum fluctuation is frozen 
when the fluctuation scale $a/k$ gets longer than 
the Hubble horizon scale $1/H$.
In LQC we will see below that the behavior of quantum fluctuation 
may be controlled by the effective horizon $\sqrt{SD}/H$ 
instead of the Hubble horizon $1/H$.

To get insight into the physical properties of the 
effective horizon we consider a massless scalar field. 
The field equations for the massless scalar field are given by 
putting $V=0$ into the equations in section II A.
The equation of motion is then given by 
	\begin{equation}
 	 \ddot{\phi}+(3-n)H\dot{\phi}=0\label{yy0}.
	\end{equation}
The above can be easily integrated to give
	\begin{equation}
	 \dot{\phi}=Ca^{n-3}\label{massless scalar},
	\end{equation}
where $C$ is an integral constant.
Substituting Eq.~(\ref{massless scalar}) into 
the Friedmann equation (\ref{f0}) with the massless scalar field, and integrating it with respect to $t$, 
we can obtain the following scale factor:
	\begin{equation}
	 a=\left[\frac{\{(r+n)-6\}C}{2}\sqrt{\frac{4\pi G}{3}\frac{S_*}{D_*}}(-t)\right]^{-\frac{2}{(n+r)-6}}+C_1,\label{scalepara00}
	\end{equation}
where $C_1$ is an integral constant, and
it should be noted that 
the scale factor increases with time for $-\infty<t<0$.

We here consider the following perturbation for the scalar field:
	\begin{equation}
	 \phi=\phi(t)+\delta\phi(t,x)\label{perturbation}.
	\end{equation}
Using Eq.~(\ref{materhamiltonian}), we can obtain the equation for 
the perturbation of the massless scalar field as
	\begin{equation}
	 \delta\ddot{\phi}+(3-n)H\delta\dot{\phi}-\frac{DS}{a^2}\nabla^2 \delta\phi=0.\label{massless00}
	\end{equation}
Here, using the Fourier transformation
	\begin{equation}
	 \delta{\phi}=\sum_k\delta\phi_{k}\exp(ikx)\label{fourier11},
	\end{equation}
and substituting Eq.~(\ref{fourier11}) into Eq.~(\ref{massless00}), 
we obtain
	\begin{equation}
	 \delta \ddot{\phi}_{k}+(3-n)H\delta \dot{\phi}_{k}+\frac{DSk^2}{a^2} \delta\phi_{k}=0\label{yy1}.
	\end{equation}
It should be noted that 
since we observe the density perturbation as a functional of the Fourier mode $\delta \phi_k$,  
from Eq.~(\ref{fooo}) the density perturbation of the massless scalar field is given by
	\begin{equation}
	 \frac{\delta\rho_k}{\rho}\simeq \frac{\delta\dot{\phi}_k}{\dot{\phi}}. \label{densityp}
	\end{equation}

First, we consider the short-wave-length limit.
Assuming $\delta\phi_{k}=\exp{(i\omega t)}$
and inserting this into Eq.~(\ref{yy1}), 
then it becomes 
	\begin{equation}
	 -\omega^2+i(3-n)H\omega+\frac{DSk^2}{a^2}=0\label{y00}.
	\end{equation}
We compare three terms on the left-hand side of Eq.~(\ref{y00}).
If 
	\begin{equation}
	 \omega H\ll\frac{DSk^2}{a^2}\label{yy02}
	\end{equation}
is satisfied, the second term is much smaller than the third term. In
that case, using Eqs.~(\ref{y00}) and (\ref{yy02}), 
we can take $\omega$ as follows
	\begin{equation}
	 \omega=\frac{\sqrt{DS}k}{a},\label{yy22}
	\end{equation}
and substituting this into Eq.~(\ref{yy02}), we obtain
	\begin{equation}
	 1\ll\frac{\sqrt{DS}k}{aH}\label{sb}.
	\end{equation}
Using Eqs.~(\ref{massless scalar}) and (\ref{scalepara00}), and $\delta\phi_{k}=\exp{(i\omega t)}$, 
we can obtain the density perturbation in the short-wave-length limit 
for the massless scalar field as follows:
	\begin{eqnarray}
	\frac{\delta\dot{\phi}_k}{\dot{\phi}} =\frac{i\omega}{C}\frac{\delta\phi_k}{a^{n-3}}
	 \propto  e^{i\omega t}(-t)^{2},
	\end{eqnarray}
where we put $r=3$. 
Notice that the density perturbation decreases with time 
in the limit of Eq.~(\ref{sb}). 

Next, we consider the long-wave-length limit and assume that the third term 
is much smaller than the other terms on the left-hand side of Eq.~(\ref{yy1}).
Then we obtain the following relation:
	\begin{equation}
	 (3-n)H\delta\dot{\phi}_{k}\gg\frac{DSk^2}{a^2} \delta\phi_k\label{yy3}.
	\end{equation}
In this case, Eq.~(\ref{yy1}) implies 
$\delta\dot{\phi}_k/\delta\phi_k\sim H$, and hence
we can rewrite Eq.~(\ref{yy3}) as follows:
	\begin{equation}
	 1\gg\frac{\sqrt{DS}k}{aH}\label{super horizon1},
	\end{equation}
where we have neglected the constant $(3-n)$.
In the above limit, 
we can solve Eq.~(\ref{yy1}) and then obtain $\delta\dot{\phi}_{k}$ as follows:
	\begin{equation}
	 \delta\dot{\phi}_{k}=C_2a^{n-3}\label{su},
	\end{equation}
where $C_2$ is an integral constant.
Using this $\delta\dot{\phi}_k$ and Eq.~(\ref{massless scalar}), 
we can obtain the following density perturbation 
in the long-wave-length limit for the massless scalar 
field:
	\begin{equation}
 	 \frac{\delta\dot{\phi}_k}{\dot{\phi}}\propto\frac{C_2}{C}=\mathrm{const}.
	\end{equation}
The density perturbation is constant in the region of Eq.~(\ref{super horizon1}).

The above argument means that 
the behavior of the density perturbation 
strongly depends on whether its length scale is
larger or smaller than $\sqrt{DS}/H$.
Thus $\sqrt{DS}/H$ acts an effectively horizon for 
the super-inflation in LQC. 
From now on, we call $\sqrt{DS}/H$ the effective horizon scale, 
and the scales satisfying Eqs.~(\ref{sb}) and (\ref{super horizon1}) are
said to be sub-horizon and super-horizon, respectively.
\subsection{The evolution of the fluctuation scale and the effective horizon scale}\label{effho}
In the super-inflation, the scale factor is given by
Eq.~(\ref{acalefactor}) on the scaling solutions.
Using this, 
the fluctuation scale $a/k$, the effective horizon scale $\sqrt{DS}/H$ 
and the Hubble horizon scale $1/H$ behave as
	\begin{equation}
	 \frac{a}{k}\propto (-\tau)^{p},\quad \frac{\sqrt{DS}}{H}\propto (-\tau)^{\frac{n+r+2}{2}p+1},\quad \frac{1}{H}\propto (-\tau)^{p+1}\label{horizons},
	\end{equation}
respectively.
We can see that the fluctuation scale gets longer than the Hubble 
horizon scale as time proceeds.
For $p>-2/(n+r)$ the fluctuation scale becomes 
longer than the effective horizon scale, while for $p<-2/(n+r)$ 
the fluctuation scale gets shorter than the effective horizon 
scale.
In our scenario we assume that the scale of fluctuation gets 
longer than the effective horizon scale to become classical
through some decoherence processes. Then we
impose the following condition on $p$:
	\begin{equation}
	 p>-\frac{2}{n+r}.\label{effectiveho}
	\end{equation}
We call this condition the super-horizon condition.
\subsection{The power spectrum in LQC}
We calculate the power spectrum in the same way in Ref~\cite{M1}.
However, note that we also incorporate the correction factor $S(q)$ into 
the matter Hamiltonian as seen in Eq.~(\ref{materhamiltonian}).
We put the scalar field perturbation given by Eq.~(\ref{perturbation}) into 
Eq.~(\ref{materhamiltonian}) as
	\begin{equation}
	 \mathcal{H}_{\phi+\delta\phi}=\frac{1}{2}\frac{D(q)p^2_{\phi+\delta\phi}}{a^3}+\frac{1}{2}aS(q)\delta^{ab}\partial_a(\phi+\delta\phi)\partial_b(\phi+\delta\phi)+a^3V(\phi+\delta\phi),\label{perturbereq}
	\end{equation}
where $p_{\phi+\delta\phi}=a^3(\dot{\phi}+\delta\dot{\phi}) /D$.
Based on the above Hamiltonian, we can calculate the power spectrum ${\mathcal P}_{\delta\phi}$ 
of the density perturbation induced 
by quantum fluctuation of the scalar field.
In particular, 
the spectral index $n_{s}$ for the density perturbation or the tilt
$b$, which is defined~\cite{L} by
\begin{equation}
b\equiv n_{s}-1\equiv \frac{d{\mathcal P}_{\delta \phi}}{dk},
\end{equation}
is a crucial observable to see the consistency between 
the prediction of the theory and the observation.
The result is given by
	\begin{equation}
	 b=3-\frac{2\sqrt{9-(6-4n-3r)2p-(12+4n-2nr-r^2-2n^2)p^2}}{2+(n+r)p}\label{d202},
	\end{equation}
where we can see the scale-invariant power spectrum is achieved by $p=0$, 
corresponding to $\beta\to \infty$
with $r=3$ in Eq.~(\ref{poten0}).
The detailed calculation is described in Appendix A.
\section{The consistency with the WMAP 5-year data}
\subsection{The potential of the scalar field}
According to the WMAP 5-year data, the CMB power spectral index is
within the range \cite{k}
	\begin{equation}
	 n_s=0.963^{+0.014}_{-0.015}.
	\end{equation}
Hence we use the following range of the tilt: 
	\begin{equation}
	 b=-0.037^{+0.014}_{-0.015}\label{wmap5}.
	\end{equation}

To constrain the power index $\beta$ of the potential from 
the observational data we use the relation between 
$\beta$ and the observed tilt $b$.
Then, with $r=3$, we solve Eq.~(\ref{d202}) for $p$ and obtain
	\begin{eqnarray}
	 p&=&\frac{2\sqrt{\zeta}-\xi}{\chi}\label{ppp1},
	\end{eqnarray}
where
	\begin{eqnarray}
	 \zeta&=&(b^2-6b+1)n^2+(2b^2-12b+42)n+33b^2-198b+441,\label{zetan} \\ 
	 \xi&=&(2b^2-12b+2)n+6b^2-36b+42, \label{xin} \\ 
	 \chi&=&(b^2-6b+1)n^2+(6b^2-36b+46)n+9b^2-54b+93. \label{chin}
	\end{eqnarray}
We can see that only this root satisfies the super-horizon 
condition (\ref{effectiveho}) and hence we have discarded another one. 
Besides, it should be noted that 
the effective horizon scale decreases with time.

We solve Eq.~(\ref{poten0}) for $\beta$ and obtain
	\begin{equation}
	 \beta=-\frac{2(2+5p)}{(n-3)p}\label{potpote0}.
	\end{equation}
Substituting Eq.~(\ref{ppp1}) into Eq.~(\ref{potpote0}) 
and using Eqs. (\ref{wmap5})--(\ref{chin}), we plot the allowed region for $\beta$ against
$n$ in FIG. 1.
If we can specify the value for $n$, then we can place a rather
stringent constraint on $\beta$.
For example, if we put $n=100$, then $\beta=25.2$. 
However, even if we do not have any information about $n$, 
$\beta$ is weakly constrained to $15\lesssim \beta\lesssim 80$.

In any case, $\beta$ is bounded from above in the present scenario and
an infinitely large value is not allowed from the WMAP 5-year data. 
This result is not so sensitive even if we slightly
expand the range of $b$ around the observed one (\ref{wmap5}).
Actually, we can see that the dependence of $\beta $ on $b$
is not monotonic. For the observed range (\ref{wmap5}) of $b$, 
$\beta$ decreases as $b$ increases. However,
as $b$ increases further to 0, $\beta$ turns to increase to infinity.
This behavior is common at least for all $n$ in the range
shown in FIG. 1. 
	\begin{figure}[h!]
        \begin{center}
              \resizebox{110mm}{!}{\includegraphics{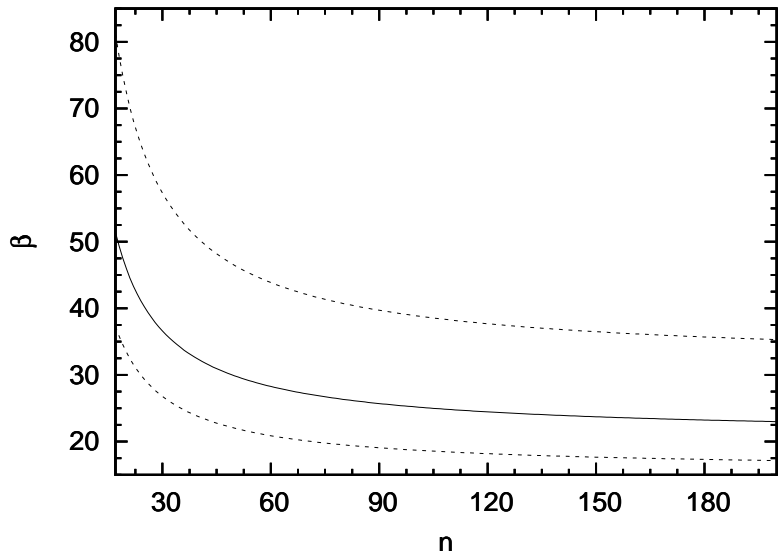}}
	      \caption{The allowed region of $\beta$ against $n$.
              The solid line corresponds to the best-fit value, while 
              the dashed lines denote the boundary of 
              measurement error ranges.}
        \end{center}\label{figeff}
	  \end{figure}

\subsection{The energy density at the end of the super-inflation}
In LQG, there is the smallest area 
element $\Delta=2\sqrt{3}\pi\gamma l_{\mathrm{Pl}}^2$~\cite{c0}.
Since the Hubble parameter increases with time, the Hubble horizon scale $1/H$ decreases and the energy density increases.
Thus, if the Hubble horizon characterizes causality and the nature
of quantum fluctuation, 
we need to show that the Hubble horizon scale is never 
smaller than the square root of the smallest area element
to validate the present analysis.
However, in the super-inflation scenario, 
causality and the nature 
of quantum fluctuation may be controlled by the effective horizon $\sqrt{SD}/H$ 
and it decreases as time proceeds.
Therefore, we need to find the condition that 
the effective horizon scale is never smaller than the square root of the 
smallest area element for the super-inflation in LQC.

In calculating the energy density at the end of the super-inflation, 
we assume the following super-inflation scenario: 
the super-inflation starts when the quantum state of 
the scalar field emerges into the classical regime~\cite{tsuji,r2} (we call this time $\tau_{\mathrm{start}}$), 
and this ends when the effective horizon becomes the Hubble horizon $\sqrt{SD}/H=1/H$ (we call this time $\tau_{\mathrm{end}}$)
and before the effective horizon scale gets
shorter than the square root of the smallest area element (we call this time $\tau_{\mathrm{Pl}}$).

We calculate the energy density from Eq.~(\ref{f0}) as
        \begin{equation}
	 \rho=\frac{3}{8\pi G S}\frac{(-p)^2}{A^2}(-\tau)^{-2(p+1)}\label{energydens00},
	\end{equation}
where we have used 
Eqs.~(\ref{s0}), (\ref{sstar}) and (\ref{acalefactor}) with $r=3$, and
	\begin{equation}
	 a_*=\tilde{a}_*l_{\mathrm{Pl}},\quad A=\tilde{A}_{*}f^{\frac{p}{2}}a_*,\quad \dot{a}=\frac{-p}{-\tau},\label{anoteigi}
	\end{equation}
where $\tilde{a}_{*}$ and $\tilde{A}_{*}$ are calculated as
	\begin{eqnarray}
	 \tilde{a}_{*}&=&\left(\frac{2j}{K}\right)^{\frac{1}{3}}\sqrt{\frac{4\pi\gamma}{3}}, \\
	 \tilde{A}_{*}&=&\left[-\frac{1}{p}\sqrt{\frac{8\pi \tilde{S}_*}{3(x_0^2-1)}}\left|\frac{2x_0}{n-r}\sqrt{\frac{3\tilde{D}_*}{4\pi \tilde{S}_*}}\right|^{\frac{\beta}{2}}\right]^p\left[\left(\frac{2j}{K}\right)^{\frac{1}{3}}\sqrt{\frac{4\pi\gamma}{3}}\right]^p.
	\end{eqnarray}
From Eqs.~(\ref{wmap5})--(\ref{chin}), 
we can show $p+1>0$ and 
see the energy density given by Eq.~(\ref{energydens00}) is 
monotonically increasing with time.
Incidentally if we take $p \rightarrow  0$ in Eq.~(\ref{energydens00}), $\rho$ approaches to a zero at the all time. 

The energy density at the end of the super-inflation is
        \begin{equation}
	 \rho_{\mathrm{end}}=\frac{3}{8\pi G}\frac{(-p)^2}{a_*^2}\tilde{S}_*^{-\frac{n}{n+3}}\tilde{D}_*^{\frac{3}{n+3}}\left(\tilde{A}_*\right)^{\frac{2}{p}}\left(\tilde{S}_*\tilde{D}_*\right)^{\frac{2(p+1)}{(n+3)p}}f,\label{wmap5obs}
	\end{equation}
where we have used Eq.~(\ref{energydens00}) and the ending time
        \begin{equation}
	 \left(-\tau_{\mathrm{end}}\right)=\left(\tilde{A}_*f^{\frac{p}{2}}\right)^{-\frac{1}{p}}\left(\tilde{S}_*\tilde{D}_*\right)^{-\frac{1}{(n+3)p}}\label{taustar},
	\end{equation}
where we have used Eqs.~(\ref{anoteigi}) and (\ref{acalefactor})
so that the effective horizon is equal to the Hubble horizon
at $\tau=\tau_{\mathrm{end}}$.
\subsubsection{the energy density at the end of the super-inflation}

Since we have the temperature fluctuation amplitude as 
$\delta \rho_k/\rho\sim \delta\dot{\phi}_k/\dot{\phi}\sim 10^{-5}$ 
in the WMAP 5-year data, 
using the amplitude and the spectral index, we will obtain 
the information about 
the energy density at the end of the super-inflation.
For this purpose, we calculate $\delta\dot{\phi}_k$ and 
$\dot{\phi}$ in terms of $f$. From Eqs.~(\ref{correctionterm01}), (\ref{phiphi}), and (\ref{acalefactor}), 
$\dot{\phi}$ is given by
	\begin{equation}
	 \dot{\phi}=\frac{(-p)x_0}{(-\tau)a}\sqrt{\frac{3D}{4\pi GS}}\label{dotphi0}
	\end{equation}
and $\delta\dot{\phi}_{k}$ can be 
calculated from the power spectrum as Eq.~(\ref{tsukaerune})
	\begin{equation}
	 <0|(\delta{\phi}_k)^{2}|0>\frac{k^3}{2\pi^2}=\mathcal{P}_{\delta\phi}(k). \label{powerphi}
	\end{equation}
So far the dimension of the scale factor is $L$ 
and the wave number $k$ is dimensionless.
Here, since we would like to make the dimensionless scale factor and 
the wave number of dimension $L^{-1}$, 
we define
	\begin{equation}
	 \tilde{a}=\frac{a}{a_*},\quad \tilde{k}=\frac{k}{a_*}, \label{neevariables}
	\end{equation}
where the characteristic scale factor $a_{*}$ has dimension $L$.
Using Eqs.~(\ref{powerphi}), (\ref{neevariables}) and (\ref{skara2}), 
we calculate $\delta\dot{\phi}_k$ as 
	\begin{equation}
	 \delta\dot{\phi}_k=\sqrt{\frac{\Gamma(|\nu|)}{4\pi^2}}\left|\frac{p}{2+(n+3)p}\right|^{\frac{1-2|\nu|}{2}}\frac{(-p)(-q)}{S^{3/4}D^{1/4}\tilde{a}^2a_*^2}\left(\frac{\sqrt{SD}a_*\tilde{k}}{(-p)}\right)^{\frac{3-2|\nu|}{2}}(-\tau)^{-\frac{1+2|\nu|}{2}}\label{dotphi01}, 
	\end{equation}
where 
	\begin{equation}
	 q=-\frac{\{2+3|\nu|-n(1-|\nu|)\}p-1+2|\nu|}{2}.
	\end{equation}
From Eqs.~(\ref{densityp}), (\ref{dotphi0}), and (\ref{dotphi01}), 
we can obtain the fluctuation of the energy density as
	\begin{equation}
	 \frac{\delta\rho_k}{\rho}=\frac{(-q)}{\tilde{a}a_*x_0(SD)^{3/4}}\sqrt{\frac{GS\Gamma(|\nu|)}{3\pi}}\left|\frac{p}{2+(n+3)p}\right|^{\frac{1-2|\nu|}{2}}\left(\frac{\sqrt{SD}a_*\tilde{k}}{(-p)}\right)^{\frac{3-2|\nu|}{2}}(-\tau)^{\frac{1-2|\nu|}{2}}\label{sukisuki},
	\end{equation}
and we assume that the fluctuation observed by WMAP is 
created at the end time of the super-inflation ($\tau=\tau_{\mathrm{end}}$), 
	\begin{equation}
	 \left(\frac{\delta\rho_k}{\rho}\right)_{obs}=\frac{(-q)}{a_*x_0}\left(\tilde{S}_*\tilde{D}_*\right)^{\frac{1}{(n+3)}}\sqrt{\frac{G\tilde{S}_*^{\frac{n}{n+3}}\tilde{D}_*^{-\frac{3}{n+3}}\Gamma(|\nu|)}{3\pi}}\left|\frac{p}{2+(n+3)p}\right|^{\frac{1-2|\nu|}{2}}\left(\frac{a_*\tilde{k}}{(-p)}\right)^{\frac{3-2|\nu|}{2}}\tilde{A}_*^{-\frac{1-2|\nu|}{2p}}f^{-\frac{1-2|\nu|}{4}}\left(\tilde{S}_*\tilde{D}_*\right)^{-\frac{1-2|\nu|}{2(n+3)p}},\label{kakikae}
	\end{equation}
where we have used Eq.~(\ref{taustar}) at this time, 
and $(\delta\rho_k/\rho)_{obs}$ is the observational 
data of the energy density fluctuation.
Here, we rewrite Eq.~(\ref{kakikae}) as follows:
	\begin{equation}
	 f=W\tilde{A}_*^{-\frac{2}{p}}\left(\tilde{S}_*\tilde{D}_*\right)^{-\frac{2}{(n+3)p}}\label{endingsuper},
	\end{equation}
where
	\begin{equation}
	 W=\left[\left(\frac{\delta\rho_k}{\rho}\right)_{obs}^{-1}\frac{(-q)}{a_*x_0}\left(\tilde{S}_*\tilde{D}_*\right)^{\frac{1}{n+3}}\sqrt{\frac{G\Gamma(|\nu|)}{3\pi}}\sqrt{\tilde{S}_*^{\frac{n}{n+3}}\tilde{D}_*^{-\frac{3}{n+3}}}\right]^{\frac{4}{1-2|\nu|}}\left|\frac{p}{2+(n+3)p}\right|^2\left(\frac{a_*\tilde{k}}{(-p)}\right)^{\frac{2(3-2|\nu|)}{1-2|\nu|}}.
	\end{equation}
Substituting Eq.~(\ref{wmap5obs}) into Eq.~(\ref{endingsuper}), 
we obtain the energy density at the end of the super-inflation as follows :
	\begin{equation}
	 \rho_{\mathrm{end}}=\frac{3(-p)^2}{8\pi l_{\mathrm{Pl}}^4\tilde{a}_*^2}W\tilde{S}_*^{-\frac{n}{n+3}}\tilde{D}_*^{\frac{3}{n+3}}\left(\tilde{S}_*\tilde{D}_*\right)^{\frac{2}{n+3}}\label{wmap5resultenergy}.
	\end{equation}

Because the effective horizon scale must not be shorter than 
the square root of the smallest area element in the present assumption,
we need the following condition of the super-inflation in LQC:
	\begin{equation}
	 \frac{\sqrt{DS}}{H}>\sqrt{\Delta}\label{effectiveho001},
	\end{equation}
where~\cite{c0}
	\begin{equation}
	 \Delta=2\sqrt{3}\pi\gamma l_{\mathrm{Pl}}^2.
	\end{equation}
At the end of the super-inflation, we substitute Eq.~(\ref{taustar}) into Eq.~(\ref{effectiveho001}), 
and rewrite this equation as 
	\begin{equation}
	 f<\frac{a_*^2}{\Delta(-p)^2}\left(\tilde{A}_*\right)^{-\frac{2}{p}}\left(\tilde{S}\tilde{D}\right)^{-\frac{2(p+1)}{(n+3)p}}\label{effectivehorizondelta}.
	\end{equation}
By substituting Eq.~(\ref{effectivehorizondelta}) into Eq.~(\ref{wmap5obs}), 
we obtain the upper limit of the energy density at 
the end of the super-inflation,
	\begin{equation}
	 \rho_{\mathrm{end}}<\frac{3}{8\pi G\Delta}\left(\tilde{S}^{-\frac{n}{n+3}}\tilde{D}_*^{\frac{3}{n+3}}\right)\label{upperlimit}, 
	\end{equation}
and we call this the effective horizon condition.

In FIG. 2, we plot the allowed region of
$\rho_{\mathrm{end}}/\rho_{\mathrm{Pl}}$ with
$\rho_{\mathrm{Pl}}=l_{\mathrm{Pl}}^{-4}$ and $\left(\delta \rho_k /\rho\right)_{obs}=10^{-5}$ for
the different values of $n$.
At least for the region shown in this figure, we can see that
the effective horizon condition (\ref{upperlimit}) is
well satisfied. Moreover, the energy scale $\rho_{\mathrm{end}}$
is far beyond that for the nucleosynthesis constraint.
In this region the energy density is much smaller than the Planck 
energy density. 
We can see that if we know the value of $n$, $\rho_{\mathrm{end}}$
is strongly constrained from the observational data. 
For example, if we put $n=100$, then $\rho_{\mathrm{end}}/\rho_{\mathrm{Pl}}=10^{-22.6}$. 
	  \begin{figure}[h!]
           \begin{center}
 	     \resizebox{110mm}{!}{\includegraphics{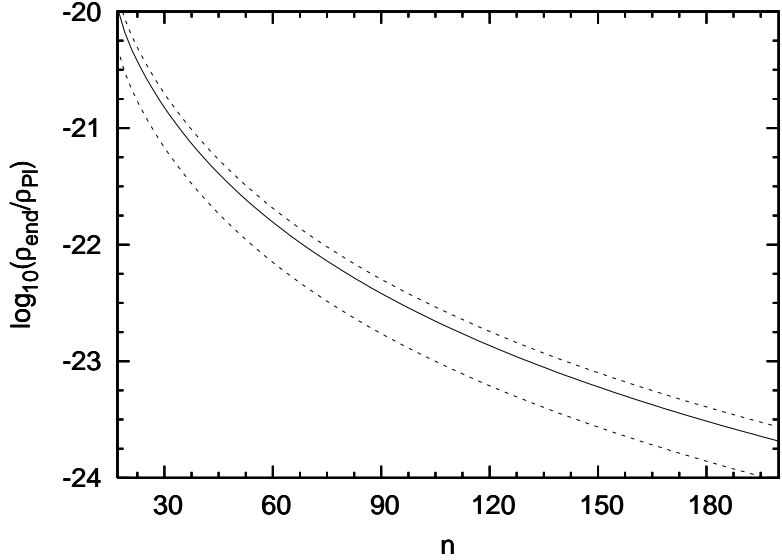}} 
           \caption{The energy density at the end of the super-inflation 
		from the WMAP 5-year data for the amplitude and the spectral index of 
		density perturbation.
           	The solid line corresponds to the best-fit value of the 
		spectral index data, while the dashed lines denote the boundary of 
		the allowed region.
           	Here, while we use value of SU(2) parameter
	    	$j=100,10^{10}$, 
		there is almost no dependence on $j$.}
	   \end{center}
          \end{figure}
\subsubsection{The period of the super-inflation}
Since the effective horizon decreases with time, 
the super-inflation must end before the effective horizon scale gets 
equal to the square root of the smallest area element to justify the present calculation.
Thus we have the following condition:
	\begin{equation}
	 \tau_{\mathrm{start}}< \tau_{\mathrm{end}}<\tau_{\mathrm{Pl}}.\label{theperiod00}
	\end{equation}
Substituting Eq.~(\ref{taustar}) into Eq.~(\ref{theperiod00}), and using Eq.~(\ref{defM}), 
we can obtain the following condition for the mass scale of the scalar field  $M$:
	\begin{equation}
	 \left[(-\tau_{\mathrm{start}})\tilde{A}_{*}^{\frac{1}{p}}\left(\tilde{S}_*\tilde{D}_*\right)^{\frac{1}{(n+3)p}}\right]^{\frac{2}{\beta-4}}> \left(\frac{M}{m_{\mathrm{Pl}}}\right)>\left[(-\tau_{\mathrm{Pl}})\tilde{A}_{*}^{\frac{1}{p}}\left(\tilde{S}_*\tilde{D}_*\right)^{\frac{1}{(n+3)p}}\right]^{\frac{2}{\beta-4}}\label{fnojouken},
	\end{equation}
where we assume $M>0$ and $15\lesssim \beta\lesssim 80$.
Here, substituting Eq.~(\ref{endingsuper}) into Eq.~(\ref{defM}), we can take the scalar field mass scale with WMAP data.
	\begin{equation}
         \left(\frac{M}{m_{Pl}}\right)=\left[W\tilde{A}_*^{\frac{1}{p}}\left(\tilde{S}_*\tilde{D}_*\right)^{\frac{1}{(n+3)p}}\right]^{\frac{2}{\beta-4}}\label{Mdata}
	\end{equation}
Since the super-inflation starts when the quantum state of 
the scalar field emerges into the classical regime, 
we can require the uncertainty principle of the scalar field
	\begin{equation}
	 \left|\phi\cdot p_{\phi}\right|> 1 \label{unsertaitnty00}
	\end{equation}
at $\tau=\tau_{\mathrm{start}}$.
Substituting Eqs.~(\ref{phiphi}) and (\ref{conjugatemoment}) into Eq.~(\ref{unsertaitnty00}), and using Eq.~(\ref{acalefactor}),
we have 
        \begin{equation}
	 \left|\frac{3x_0^2a^2}{2(n-3)\pi GS}\frac{(-p)}{(-\tau)}\right|>1\label{unsertainty001}.
	\end{equation}
By substituting Eq.~(\ref{endingsuper}) into Eqs.~(\ref{unsertainty001}) and (\ref{effectiveho001}),
we can determine $(-\tau_{\mathrm{start}})$ and $(-\tau_{\mathrm{Pl}})$ as
	\begin{equation}
	 (-\tau_{\mathrm{start}})>\left[\frac{3x_0^2\tilde{a}_*^2(-p)}{2(n-3)\pi\tilde{S}_*}\right]^{\frac{1}{p+1}}W^{-\frac{p}{2(p+1)}}\left(\tilde{S}_*\tilde{D}_*\right)^{\frac{1}{(n+3)(p+1)}}\label{unsertau},
	\end{equation}
	\begin{equation}
	 (-\tau_{\mathrm{Pl}})=\left[\sqrt{\frac{\Delta}{\tilde{S}_*\tilde{D}_*}}\frac{(-p)}{a_*}\right]^{\frac{2}{(n+5)p+2}}W^{-\frac{(n+5)p}{2(2+(n+5)p)}}\left(\tilde{S}_*\tilde{D}_*\right)^{\frac{(n+5)}{2(2+(n+5)p)(n+3)}}.\label{effho002}
	\end{equation}
It turns out that the uncertainty principle condition does not 
essentially constrain the parameters of the scenario.
Then, we substitute Eq.~(\ref{effho002}) into 
Eq.~(\ref{fnojouken}), which gives a lower bound of the scalar field
mass scale $M$. In FIG. 3, we plot the allowed region from the 
observational constraint~(\ref{Mdata}) with $b=-0.037$ and $\left(\delta \rho_k/\rho\right)_{obs}=10^{-5}$
for the different values of $n$.
At least in the region shown in FIG. 3, we can easily show that
the allowed region is far beyond the lower bound in 
Eq.~(\ref{fnojouken}) and the scalar filed mass scale has the Planck mass order.
If we can know the volume correction parameter $n$,
then we can constrain the scalar field mass scale $M$ rather stringently. 
For example, if we put $n=100$, then $M/m_{\mathrm{Pl}}=0.150$.
	  \begin{figure}[h!]
           \begin{center}
 	     \resizebox{110mm}{!}{\includegraphics{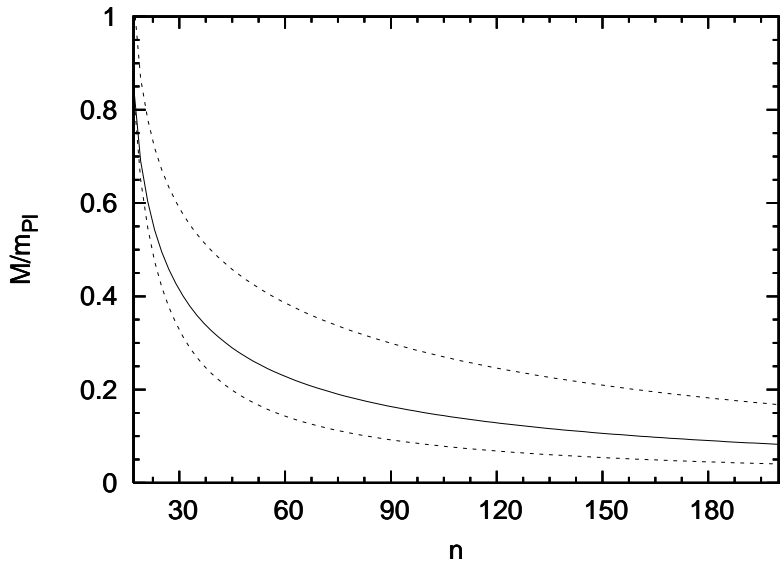}} 
           \caption{The allowed region of the scalar field mass scale $M$ 
		constrained from the observational data for the 
		density perturbation. The solid line denotes the best-fit value, 
		while the dashed lines denote the boundary of the allowed region.
		There is almost no dependence on the value 
            	of SU(2) parameter $j$ for
	    	$j=100,10^{10}$.}
	   \end{center}
          \end{figure}

\section{Conclusion}
We have considered a single scalar field with the negative power-law potential $V=-M^4\left(\phi/M\right)^\beta$ in LQC, 
and determined the allowed region of the potential power index $\beta$ and the energy density at the end of the super-inflation $\rho_{\mathrm{end}}$, 
and the scalar field mass scale $M$ by using the consistency with the WMAP 5-year data. 

First we have reviewed Ref~\cite{M2}. Using dynamical systems theory we
have found scaling solutions which are stable fixed points 
and satisfy the super-horizon condition. 
Second, we have determined the super-horizon 
condition~(\ref{effectiveho}) by the behavior of the density
perturbation of a massless scalar field, 
and then considered the effective horizon scale instead of the Hubble horizon scale. 
Third, we have assumed the super-inflation scenario: 
the super-inflation starts when the quantum state of the scalar field 
emerges into the classical regime
and ends when the effective horizon gets equal to the Hubble horizon 
before the effective horizon scale gets shorter 
than the square root of the smallest area element.
Finally, by using the above inflation scenario and the consistency with
the WMAP 5-year data, we have reached the following conclusion.
If we can specify the volume correction parameter $n$, 
we can constrain the potential parameters $\beta $ and $M$, and the energy density at the end of the super-inflation $\rho_{end}$ rather 
stringently.
Even if we only know the volume correction parameter $n$ in the range as
$81/5<n<\infty$, we can constrain $\beta$, $M$ and $\rho_{\mathrm{end}}$ as follows: 
$\beta$ exists in the region as $15\lesssim \beta\lesssim 80$, $M$ is the order of the Planck mass, 
and $\rho_{\mathrm{end}}$ is smaller than the Planck energy density, respectively. 
Besides, for example, if we put $n=100$, we can constrain $\beta$, $M$ and $\rho_{\mathrm{end}}$ as follows:
$\beta=25.2$, $\rho_{\mathrm{end}}/\rho_{\mathrm{Pl}}=10^{-22.6}$, and $M/m_{\mathrm{Pl}}=0.150$.
The reason why we have obtained the upper bound on the power index
$\beta$ in contrast to the previous works~\cite{M1,M2} is that we have 
considered the observed spectral index for 
the CMB power spectrum in the WMAP 5-year data,
which significantly favors a red power spectrum.   

In this paper we have only considered the scalar field perturbation
and directly related it to the density perturbation, 
and have assumed the negative potential only in the super-inflation.
Since the semiclassical LQG effects will become insignificant as the universe expands, 
the super inflation lasts only for a finite interval of time. After that the 
motion of the scalar field will be irrelevant and the potential will 
not be described by the negative potential, 
so the super-inflation might turn to the standard chaotic inflation when it ends. 
We would simply assume that the scalar field perturbation amplitude
may be of the same order as the temperature perturbation in
the observed CMB anisotropy. 
However, the observed CMB anisotropy is actually the temperature perturbation
on the last scattering surface, so we need the consistent
formulation of matter and curvature perturbations in LQC with the appropriate
treatment of gauge freedom and the detailed analysis of their evolution in different scales.
To formulate the consistent perturbation formulation, Bojowald et al.~\cite{Bojowald:20080, Bojowald:2008}
recently indicated that anomaly cancellation should occur in the
effective theory and this strongly restricts the possible effective theory.
From these point of view, our analysis here considered is a toy model which
does not take into account backreaction and anomaly cancellation.
Therefore, the present assumption that the density perturbation of the
scalar field in this simplified framework is directly comparable with
the observed CMB power spectrum at least in order of magnitude is to be
under careful investigation.
To get more robust constraint on the super-inflation scenario, 
we will need to use other independent observations: 
large scale structure, non-Gaussian, gravitational waves (cf.~\cite{M3}) and so on. 
These problems will be our next work.
As mentioned in section $\mathrm{II}$,
there is an open problem in introducing the characteristic scale factor $a_*$
into the flat FRW universe, in which the overall constant factor in the scale
factor is just a gauge freedom. Although this problem deserves careful attention
and more work is required to clarify the validity of the dynamical equations,
we have chosen here to proceed by demonstrating a method of obtaining the
observational constraint in the present scenario, which can easily be employed
once progress is made on this currently uncertain section of the theory.
See Ref~\cite{G.Calcagni1, G.Calcagni2} for a recent interesting attempt to resolve this important issue.
\section*{Acknowledgment}
We are grateful to R. Tavakol for several discussions, and thankful to A. Ishibashi for helpful comment. 
MS is supported by Rikkyo University Special Fund for Research, and 
TH was partly supported by the Grant-in-Aid for Scientific Research Fund of the Ministry of Education, 
Culture, Sports, Science and Technology, Japan (Young Scientists (B) 18740144).
\newpage
\begin{appendix}
\section{The power spectrum in LQC}\label{appendA}
We calculate the power spectrum of quantum fluctuation 
based on the Hamiltonian~(\ref{perturbereq}).
We can obtain the equation of motion for the perturbation as
	\begin{equation}
	 \delta\phi''=\left[-2\frac{a'}{a}+\frac{D'}{D}\right]\delta\phi'+D\left[S\nabla^2 -a^2\frac{d^2V}{d\phi^2}\right]\delta\phi. \label{eq62}
	\end{equation}
Now, we use the following variables in Eq.~(\ref{eq62}):
	\begin{eqnarray}
	 u&=&\frac{a}{\sqrt{D}}\delta\phi\label{skara0}, \\
	 m_{\mathrm{eff}}^2&=&-\left(\frac{a}{\sqrt{D}}\right)''\frac{\sqrt{D}}{a}+a^2D\frac{\partial^2V}{\partial\phi^2}\label{meff},
	\end{eqnarray}
and then we obtain
	\begin{equation}
	 u''+\left(-SD\nabla^2+m_{\mathrm{eff}}^2\right)u=0\label{sks0}.
	\end{equation}
Besides, we obtain a plane wave expansion of $u$ as
	\begin{equation}
	 u=\int\frac{d^3k}{(2\pi)^{3/2}}\left[w_k(\tau)\hat{a}_ke^{ikx}+w_k^*(\tau)\hat{a}_k^{\dag}e^{-ikx}\right],\label{eq66}
	\end{equation}
where $\hat{a}_{k}^{\dag}$ and $\hat{a}_{k}$ are creation and annihilation operators, which satisfy the usual commutation relations 
	\begin{equation}
	 \left[\hat{a}_k,\hat{a}_{k'}^{\dag}\right]=\delta(k-k'),\quad \left[\hat{a}_k,\hat{a}_{k'}\right]=\left[\hat{a}_k^\dag,\hat{a}_{k'}^\dag\right]=0.
	\end{equation}
Substituting Eq.~(\ref{eq66}) into Eq.~(\ref{sks0}), we have
	\begin{equation}
	 w_k''+\left(SDk^2+m_{\mathrm{eff}}^2\right)w_k=0 \label{w0}.
	\end{equation}
The canonical quantization for $w_k$ and its conjugate momentum requires
	\begin{equation}
	 w_k^*\frac{dw_k}{d\tau}-w_k\frac{dw_k^*}{d\tau}=-i\label{wronski}. 
	\end{equation}
We use the Fourier transformation of $\tilde{u}_k(\tau)$ as 
	\begin{equation}
	 \tilde{u}_k(\tau)=\int d^3xe^{-ikx}u, 
	\end{equation}
and then the power spectrum $\mathcal{P}_u$ of quantum fluctuation
	\begin{equation}
	 \langle 0|\tilde{u}_k^\dag\tilde{u}_k|0\rangle
=\frac{2\pi^2}{k^3}\mathcal{P}_u(k)\label{tsukaerune}
	\end{equation}
is given by 
	\begin{equation}
	 \mathcal{P}_u(k)=\frac{k^3}{2\pi^2}|w_k|^2\label{eqmo},
	\end{equation}
where we have defined the vacuum as $\hat{a}_k|0\rangle=0$.

We need to solve Eq.~(\ref{w0}) to calculate 
the right-hand side of Eq.~(\ref{eqmo}).
We use the following variables: 
	\begin{eqnarray}
	 \nu&=&-\frac{\sqrt{1-4m_{\mathrm{eff}}^2\tau^2}}{2+(n+r)p}\label{nu},\\
	 \psi&=&\alpha k(-\tau)^{\{2+(n+r)p\}/2}=\Bigl|\frac{2p}{2+(n+r)p}\Bigr|\frac{\sqrt{SD}k}{aH}\label{x3},
	\end{eqnarray}
where we have used $\alpha=(2\sqrt{S_*D_*A^{n+3}})/|2+(n+r)p|$.
Using these variables, the left-hand side of Eq.~(\ref{w0}) is rewritten as
	\begin{eqnarray}
	w_k''+\left(SDk^2+m_{eff}^2\right)w_k
	  =  \frac{d^2w_k}{d\psi^2}+\left[\frac{1}{\psi}+\frac{1}{\sqrt{SD}k\tau}\right]\frac{dw_k}{d\psi}+\left[1+\frac{1}{4\tau^2SDk^2}-\frac{\nu^2}{\psi^2}\right]w_k,
	\end{eqnarray}
and then the solution of Eq.~(\ref{w0}) is given by
	\begin{equation}
         w_k(-\tau)=\sqrt{\frac{\pi}{2|2+(n+r)p|}}\left\{d_1\sqrt{-\tau}H_{|\nu|}^{(1)}(\psi)+d_2\sqrt{-\tau}H_{|\nu|}^{(2)}(\psi)\right\}\label{skara1},
        \end{equation}
where $d_1$ and $d_2$ are constants and have the 
relation $|d_1|^2-|d_2|^2=1$ to satisfy Eq (\ref{wronski}).
$H_{|\nu|}^{(1)}(\psi)$ and $H_{|\nu|}^{(2)}(\psi)$ are the Hankel functions which 
are given by $H_{|\nu|}^{(1)}(\psi)=J_{|\nu|}(\psi)+iY_{|\nu|}(\psi)$ 
and $H_{|\nu|}^{(2)}(\psi)=J_{|\nu|}(\psi)-iY_{|\nu|}(\psi)$, respectively,
where $J_{|\nu|}(\psi)$ is the first Bessel function and $Y_{|\nu|}(\psi)$ is the second Bessel function.
We chose the mode function so that the vacuum state becomes 
the Bunch-Davis like one 
in the short-wave-length limit $\psi\gg1$ or $\sqrt{HD}/H\gg a/k$, 
(if we take $n=r=0$ at the classical region, 
then the solution corresponds to the Bunch-Davis vacuum \cite{bi}). 
Then we get
	\begin{equation}
	 w_k=\frac{(-\tau)^{\frac{-(n+r)p}{4}}}{\sqrt{|2+(n+r)p|\alpha k}}e^{[i\alpha k(-\tau)^{\{2+(n+r)p\}/2}]}\label{banci}.
	\end{equation}
In this limit, Eq.~(\ref{skara1}) becomes as follows, 
	\begin{equation}
	 w_k(-\tau)=\frac{(-\tau)^{\frac{-(n+r)p}{4}}}{\sqrt{|2+(n+r)p|\alpha k}}\left\{d_1\exp{[i\alpha k(-\tau)^{\{2+(n+r)p\}/2}]}+d_2\exp{[-i\alpha k(-\tau)^{\{2+(n+r)p\}/2}]}\right\}.
	\end{equation}
In the long-wave-length limit $\psi\ll 1$ or $\sqrt{HD}/H\ll a/k$, 
the first and second Bessel functions are
	\begin{eqnarray}
	 J_{|\nu|}&\longrightarrow& \frac{1}{\Gamma(|\nu|+1)}\left(\frac{\psi}{2} \right)^{|\nu|}, \nonumber \\
	 Y_{|\nu|}&\longrightarrow& -\frac{\Gamma(|\nu|)}{\pi}\left(\frac{\psi}{2} \right)^{-|\nu|}.
	\end{eqnarray}
Hence $H_\nu^{(1)}(\psi)\sim iY_{|\nu|}$, 
the power spectrum becomes
	\begin{eqnarray}
	 \mathcal{P}_{\delta\phi}(k)=\frac{\Gamma(|\nu|)}{4\pi^2}\left|\frac{p}{2+(n+3)p}\right|^{1-2|\nu|}\frac{H^2}{S^{3/2}D^{1/2}}\left(\frac{\sqrt{SD}k}{aH}\right)^{3-2|\nu|}
	 \propto k^{3-2|\nu|}(-\tau)^{1-|\nu|\{(n+r)p+2\}},\label{skara2}
	\end{eqnarray}
where we have used $\mathcal{P}_{\delta \phi}(k)=\mathcal{P}_{u}(k)D/a^2 $.
We define the tilt $b$ as the exponent of $k$ in the above equation~\cite{L}.
To calculate this in terms of $p$, we substitute 
Eq.~(\ref{meff}) into Eq.~(\ref{nu}) using
Eqs.~(\ref{correctionterm01}), (\ref{plow}) and (\ref{acalefactor}).
Then the tilt becomes the following:
	\begin{equation}
	 b=3-\frac{2\sqrt{9-(6-4n-3r)2p-(12+4n-2nr-r^2-2n^2)p^2}}{2+(n+r)p}\label{d2}.
	\end{equation}
\end{appendix}


\begin{thebibliography}{99}
        \bibitem{c0} C. Rovelli, {\it Quantum Gravity}, (Cambridge University Press, Cambridge, 2004).
        \bibitem{t2} T. Thiemann, {\it Modern Canonical Quantum General	Relativity}, (Cambridge University Press, Cambridge, 2007).
        \bibitem{Ashtekar_2005} A.~Ashtekar, New J. Phys. {\bf 7}, 198 (2005).
        \bibitem{b0} M. Bojowald, Class. Quant. Grav. {\bf 17}, 1489 (2000).
        \bibitem{b1} M. Bojowald, Class. Quant. Grav. {\bf 17}, 1509 (2000).
        \bibitem{b2} M. Bojowald, Class. Quant. Grav. {\bf 18}, 1055 (2001).
        \bibitem{b3} M. Bojowald, Class. Quant. Grav. {\bf 18}, 1071 (2001).
        \bibitem{bojo0} M. Bojowald, Phys. Rev. Lett. {\bf 86}, 5227 (2001).
	\bibitem{bojo1} M. Bojowald, Class. Quant. Grav. {\bf 19}, 5113 (2002).
        \bibitem{bojo2} M. Bojowald, Phys. Rev. Lett. {\bf 89}, 261301 (2002).
        \bibitem{Date:2004} G. Date and G. M. Hossain, Class. Quant. Grav. {\bf 21}, 4941 (2004).
        \bibitem{Banerjee:2005} K. Banerjee and G. Date, Class. Quant. Grav. {\bf 22}, 2017 (2005).
        \bibitem{Singh:2005} P. Singh and K. Vandersloot, Phys. Rev. D. {\bf 72}, 084004 (2005).
        \bibitem{As1} A. Ashtekar, T. Pawlowski and P. Singh, Phys. Rev. D {\bf 73}, 124038 (2006).
 	\bibitem{As2} A. Ashtekar, T. Pawlowski and P. Singh, Phys. Rev. D {\bf 74}, 084003 (2006).
 	\bibitem{As3} A. Ashtekar, T. Pawlowski and P. Singh, and K. Vandersloot, Phys. Rev. D {\bf 75}, 024035 (2007).
        \bibitem{tsuji} S. Tsujikawa, P. Singh, and R. Maartens, Class. Quant. Grav. {\bf 21}, 5767 (2004).
        \bibitem{Xin} X. Zhang and Y. Ling, JCAP {\bf 0708}, 012, (2007).
        \bibitem{M1} D. J. Mulryne and N. J. Nunes,  Phys. Rev. D {\bf 74}, 083507 (2006).
        \bibitem{M2} E. J. Copeland, D. J. Mulryne, N. J. Nunes, and M. Shaeri, Phys. Rev. D {\bf 77}, 023510 (2008).
        \bibitem{V1} K. Vandersloot, Phys. Rev. D {\bf 71}, 103506 (2005).
	\bibitem{k} E. Komatsu, et al.,  Astrophys. J. Suppl. {\bf 180}, 330 (2009).
        \bibitem{bojo3} A. Ashtekar, M. Bojowald, and J. Lewandowski, Adv. Theor. Math. Phys {\bf 7}, 233 (2003).
        \bibitem{tamaki_nomura_2005} T. Tamaki and H. Nomura, Phys. Rev. D {\bf 72}, 107501 (2005).
        \bibitem{s1} J. Magueijo and P. Singh, Phys. Rev. D {\bf 76}, 023510 (2007).
        \bibitem{G.Calcagni1} G. Calcagni and M. Cortes, Class. Quant. Grav. {\bf 24}, 829 (2007).
        \bibitem{G.Calcagni2} G. Calcagni and G. M. Hossain, Adv. Sci. Lett. {\bf 2}, 184 (2009).
        \bibitem{L} A. R. Liddle and D. H. Lyth, {\it Cosmological Inflation and Large-Scale Structure}, (Cambridge University Press, Cambridge, 2000).
        \bibitem{r2} M. Bojowald, J. E. Lidsey, D. J. Mulryne, P. Singh, and R. Tavakol, Phys. Rev. D {\bf 70}, 043530 (2004).
        \bibitem{Bojowald:20080}M.~Bojowald, G.~M.~Hossain, M.~Kagan, and S.~Shankaranarayanan, Phys. Rev.D {\bf 78}, 063547 (2008). 
        \bibitem{Bojowald:2008}M.~Bojowald, G.~M.~Hossain, M.~Kagan, and S.~Shankaranarayanan, Phys. Rev. D {\bf 79}, 043505 (2009).
        \bibitem{M3}  E. J. Copeland, D. J. Mulryne, N. J. Nunes, and M. Shaeri, Phys.Rev.D {\bf 79}, 023508 (2009).
        \bibitem{bi} N. D. Birrel and P. C. W. Davies, {\it Quantum fields in curved space}, (Cambridge University Press, Cambridge, 1982).
 \end{thebibliography}
\end{document}